# EFFECT OF β-DYSTROGLYCAN PROCESSING ON UTROPHIN / DP116 ANCHORAGE IN NORMAL AND MDX MOUSE SCHWANN CELL MEMBRANE

K. HNIA[a, b], G. HUGON[a], A. MASMOUDI[b], J. MERCIER[a], F. RIVIER[a] & D. MORNET[a], *

[a] EA 701, Institut de Biologie, Boulevard Henri IV, 34060 Montpellier, France

[b] Institut Supérieur de Biotechnologie & UR. 08/39 Faculté de Médecine, Monastir, Tunisia

*To whom correspondence should be addressed: Tel: +33 (0) 467 600 765. Fax: +33 (0) 467606904. E-mail address: dominique.mornet@univ-montp1.fr

Abbreviations: PN, peripheral nerve; Dp, dystrophin product; Up, utrophin product; DG, dystroglycan; DB, dystrobrevin; ITG, integrin; MMP, metalloproteinase; SDS, sodium dodecyl sulfate; PAGE, polyacrylamide gel electrophoresis; RT-PCR, reverse transriptase-polymerase chain reaction.

In the peripheral nervous system, utrophin and the short dystrophin isoform (Dp116) are co-localized at the outermost layer of the myelin sheath of nerve fibers; together with the dystroglycan complex. Dp116 is associated with multiple glycoproteins, i.e. sarcoglycans, and α- and β-dystroglycan, which anchor the cytoplasmic protein subcomplex to the extracellular basal lamina. In peripheral nerve, matrix metalloproteinase (MMP) activity disrupts the dystroglycan complex by cleaving the extracellular domain of β-dystroglycan. MMP creates a 30 kDa fragment of β-dystroglycan, leading to a disruption of the link between the extracellular matrix and the cell membrane. In this study, we investigated the molecular interactions of full length and 30 kDa β-dystroglycan with Dp116 and utrophins in peripheral nerve Schwann cells from normal and mdx mice. Our results showed that Dp116 had greater affinity to the glycosylated form of β-dystroglycan than the 30 kDa form. Interestingly, the short isoform of utrophin (Up71) was highly expressed in mdx Schwann cells compared with normal Schwann cells. In contrast to Dp116, Up71 had greater affinity to the 30 kDa β-dystroglycan. These results are discussed with regard to the participation of the short utrophin isoform and the cleaved form of β-dystroglycan in mdx Schwann cell membrane architecture and their possible role in peripheral nerve physiology.

Dystroglycan is encoded by a single gene and cleaved into two proteins by post-translational processing, resulting in an extracellular peripheral membrane glycoprotein, α-dystroglycan, and an integral membrane glycoprotein, β-dystroglycan (Ibraghimov-Beskrovnaya et al. 1992). In skeletal muscle, α-dystroglycan links laminin-2 and agrin in the basal lamina with $\alpha$-dystroglycan in the sarcolemma (Ibraghimov-Beskrovnaya et al. 1992; Ervasti and Campbell 1993; Gee et al. 1993; Fallon and Hall 1994). On the cytoplasmic side of the sarcolemma, β-dystroglycan (β-$DG_{full}$: 43 kDa) is anchored to the cytoskeletal protein dystrophin (Jung et al. 1995). In addition, β-dystroglycan interacts with Grb2, an adaptor protein, and rapsyn, a peripheral protein required for acetylcholine receptor clustering, which suggests that the dystroglycan complex may also have signalling functions in skeletal muscle (Yang et al. 1995; Cartaud et al. 1998). The dystroglycan complex is also expressed in non-muscle tissues. In peripheral nerve, α- and β-dystroglycans are expressed in the Schwann cell outer membrane apposing the endoneurial basal lamina but not in the inner membrane or compact myelin (Matsumura et al. 1993; Yamada et al. 1996). In the Schwann cell, α-dystroglycan links laminin-2 and agrin in the endoneurium with β-dystroglycan in the outer membrane (Feltri et al. 1999; Saito et al. 1999). Because peripheral myelination is greatly disturbed in congenital muscular dystrophy patients and dy mice deficient in laminin-2 (Matsumura et al. 1997), these findings suggest an involvement of the dystroglycan complex with laminin-2 in peripheral myelinogenesis (Masaki et al. 2002). Dysfunction of the dystroglycan complex has generally been implicated in the molecular pathogenesis of severe forms of hereditary neuromuscular diseases (Cohn et al. 2002; Matsumoto et al. 2005; Matsumura et al. 2005; Saito et al. 2005; van Reeuwijk et al. 2005). In this context, several studies have investigated proteolytic processing of the dystroglycan complex. A 30 kDa fragment of β-dystroglycan (β-$DG_{30}$) was detected in normal peripheral nerve, kidney, lung, and smooth muscle, but not in skeletal muscle, cardiac muscle or brain (Yamada et al. 2001; Zaccaria et al. 2001). This fragment is the product of proteolytic processing of the extracellular domain of β-dystroglycan by the membrane-associated matrix metalloproteinase (MMP) (Yamada et al. 2001). Processing disintegrates the dystroglycan complex and causes disruption of the link between the extracellular matrix (ECM) and the cell membrane, which could have profound effects on cell viability (Kaczmarek et al. 2002).

The cytoskeletal protein dystrophin and its autosomal homologue utrophin are expressed in sarcolemma, where they serve as a link between the intracellular cytoskeleton and the extracellular matrix via the dystroglycan complex (Ibraghimov-Beskrovnaya et al. 1992; Ervasti and Campbell 1993). Both proteins have also been localized in nervous tissues such as peripheral nerve or brain (Love et al. 1993; Matsumura et al. 1993). Distal transcripts of the dystrophin gene have been identified and shown to be expressed differently from the full size 427 kDa dystrophin. Dp116 is expressed in the Schwann cells of peripheral nerve (Byers et al. 1993). Utrophin, a foetal homologue of dystrophin, is localized exclusively in the neuromuscular junction in adult skeletal muscle (Nguyen et al. 1991). Utrophin also has various transcripts, as previously described (Wilson et al. 1999). A short isoform with a molecular weight of 70 kDa (Up71) was detected in peripheral nerve (Fabbrizio et al. 1995; Rivier et al. 1996). Despite these discoveries, a number of questions remain concerning the implication of either the β-$DG_{30}$ or the β-$DG_{full}$ form in the anchorage of utrophin isoforms and Dp116 in Schwann cell membrane. We questioned whether these interactions were different in normal (C57BL/10) and dystrophin-deficient mdx mouse Schwann cells. Dp116 and utrophin have a C-terminal domain containing the β-dystroglycan binding site (Byers et al. 1993; Matsumura et al. 1993). No studies have yet investigated the impact of β-dystroglycan processing on molecular interactions with full- and short-length utrophin. The short transcript Up71 is predicted to encode protein that has lost the N-terminal actin-binding domain and part of or the entire long spectrin like rod domain. Up71 retains the C-terminal domains that mediate binding to β-DG. In this study we have characterized the expression of Up71 isoform in peripheral nerve Schwann cells from normal and mdx mice. In addition we have investigated the effect of β-DG$_{full}$ processing into β-DG$_{30}$ by MMPs on molecular interaction with Dp116 and utrophin isoforms in normal and mdx Schwann cell membrane.

**EXPERIMENTAL PROCEDURES**

**Animals and sciatic nerve dissection**

Six month-old control (C57BL/10) and mdx mice were purchased from Jackson Laboratories (Bar Harbor, ME) and bred in our animal facilities. Mice were housed in plastic cages in a temperature-controlled environment with a 12-hour light/dark cycle and free access to food and water. The investigation conformed to the Guide for the Care and use of Laboratory animals published by the National institute of health (NIH). The animals were killed by rapid cervical dislocation. Animal experiments were carefully designed to minimize the number of animals and their suffering. Peripheral sciatic nerves (PN) were dissected and divided into two parts. One part was rapidly frozen in 2-methylbutane, cooled in liquid nitrogen and stored at -80°C until use for total protein extraction and immunohistochemical analysis. The second part was used as explants to make Schwann cell culture.

**Antibodies**

Polyclonal antibodies against β-dystroglycan (LG5), α-dysrobrevin1 (D124), α7B-integrin (K6) and dystrophin (H4) were produced and characterized, as previously described (Garcia-Tovar et al. 2002; Marquez et al. 2003; Royuela et al. 2003b). Polyclonal antibody against the utrophin C-terminal end was produced as follows: utrophin C-terminal residues (3423-3433) was used as a specific antigenic sequence and the K7 antibody was obtained by injecting the KLH-linked peptide as antigen, according to a previously described protocol (Royuela et al. 2003a). Monoclonal antibody (5A3) that recognized both dystrophin and utrophin was produced and characterized, as previously described (Fabbrizio et al. 1995; Rivier et al. 1996; Hernandez-Gonzalez et al. 2005). Rabbit polyclonal antibody pan-Na$^+$ and the Schwann cell marker S-100 antibody were from Chemicon International. The monoclonal antibodies directed against dystrophin (DYS2), utrophin (DRP1), and β-dystroglycan (NCL-b-DG), were purchased from Novocastra.

**Histochemistry and immunofluorescence light microscopy**

Cryostat sections (transversal 10 µm) of unfixed and dissected PN (for Nodes of Ranvier and Schwann cell visualization) were double labelled with the polyclonal antibodies described above and S-100 or pan-Na$^+$ antibodies. Immunoreactions were detected with Cy3- or FITC-conjugated sheep anti-rabbit IgG (Euromedex). All antibodies were tested in competition with corresponding synthetic peptides on cryostat sections. No immunofluorescence labelling was observed when peptides were applied. For negative controls only the second antibody was applied. Labelled sections were visualized under a Nikon optiphot-2 microscope and fluorescence intensity was analysed using Histolab program version 5-13-1 (Microvision Instrument, licence number; ).

**Total PN protein extraction and Western blotting**

First, 0.01 g of PN was homogenized in 150 µl of 5% SDS buffer (50 mM Tris/HCl, pH 8.0, 10 mM EDTA, 5% SDS) supplemented with 1% trypsin inhibitor and 1% saponin. After centrifugation (10 min at 13,000 ×*g*), protein concentration was estimated in the supernatant using the BCA protein assay kit (Pierce). Then, 50 µl of SDS buffer containing 0.01% bromophenol blue, 10% glycerol and 5% β-mercaptoethanol was added to the supernatant and each sample was denatured for 5 min at 100°C. Protein extracts were submitted in duplicate to SDS-PAGE (3-10% or 5-15%). The resulting gel was transferred onto 0.2 μm nitrocellulose membrane. Each blot was saturated in Tris-buffered saline with 0.1% Tween 20 (TBST) containing 3% bovine serum albumin (w/v). All membranes were incubated with primary antibodies for 1 h at room temperature. After labelling, the membranes were washed in TBST and then incubated with a phosphatase-labelled second antibody (Chemicon International). Antibody-bound proteins were detected with NBT/BCIP substrate, as previously described (Chazalette et al. 2005).

**Zymography**

Zymography was carried out according to Heussen and Dowdle (Heussen and Dowdle 1980). Zymogram gel (5-15% polyacrylamide) was impregnated with gelatin at 1mg/ml. After electrophoresis, the gel was washed in 2.5% Triton X100 solution at room temperature and incubated for 24 h in a substrate buffer (50mM Tris-Hcl, pH 8.0, 5mM $CaCl_2$, 0.02% $NaN_3$) at 37°C. MMPs are secreted in a latent form and require cleavage of a peptide from their $NH_2$ terminus for activation. However, exposure of proenzymes of the tissue extracts to SDS during the gel separation procedure leads to activation without proteolytic cleavage. The gel was stained in Coomassie Blue R250 for 1 h and destained in water overnight. Gelatin-degrading enzymes were visualized as clear bands, indicating proteolysis of the substrate protein. Gel was treated in black/white colour and the MMP bands were quantified using the NIH image program.

**Schwann cell culture**

The nerve explants were sterilely divided into 1-mm segments and dissociated enzymatically using 0.05% collagenase (Sigma) and 0.1% hyaluronidase in 1 ml of F-12 solution (Gibco) at 37°C for 2 h. The mixture was filtered through a 100-μm cell filter and centrifuged at 600 ×g for 5 min. The supernatant was discarded and the pellet was suspended in Dulbecco's Minimum Eagle's Medium (DMEM; Gibco) containing 10% foetal bovine serum (FBS; Sigma), 2mM L-glutamine (Gibco), 1% streptavidin (Gibco), and 1% streptomycin (Gibco). The cell suspension was plated on a 75 cm$^2$ flask (Nunc) and kept in a humidified atmosphere of 5% carbon dioxide at 37°C. Medium was replaced every 48 h. Isolation of Schwann cells was recorded according to a laminin-selection process recently described by Pannunzio et al. (Pannunzio et al. 2005). The primary culture was plated using laminin (Sigma) flasks coated immediately before use with 5µg/ml laminin solution. After passages through the laminin-selection process, cells were maintained in culture until confluent with the previously described medium supplemented with 2 µM forskolin (Sigma) to stimulate Schwann cell mitogenesis. The nature of the selected cells was determined visually based on cell morphology and Schwann cell marker S-100 by immunoflurescence. The homogeneity of the Schwann cells after five passages reached 80-90%.

**Crude membrane fraction of PN Schwann cells**

Crude membrane fraction was prepared from 5 mg of Schwann cells homogenized in a Triton X-100 lysis buffer (120 mM NaCl and 1% Triton X-100) supplemented with 10 mM iodoacetamid, 1% saponin, 1% trypsin inhibitor, 20 mM PMSF and 1% leupeptin, as previously described (Yamada et al. 1996; Saito et al. 1999; Yamada et al. 2001). Homogenates were then centrifuged (140,000 ×g) for 30 min at 4°C and the supernatant was intensively washed and adjusted at the protein concentration of 5mg/ml in the lysis buffer for immunoprecipitation experiments.

**Enzymatic deglycosylation of the endogenous β-DG$_{full}$**

Neuraminidase (BioLabs) treatment was performed according to the protocol described by Yamada et al. (Yamada et al. 1996). Briefly, purified crude membrane was diluted in SDS solution and incubated at 100°C for 5 min. It was then diluted 10-fold with concentrated buffer and distilled water to a final concentration of 0.5 M acetate buffer, pH 5.5, 1% Triton X-100, and 1% SDS, and incubated with neuraminidase (10 mU/μg of substrate) at 37°C for 24 h. For control, samples were treated identically in the absence of neuraminidase. The deglycosylation reaction was stopped by adding SDS sample buffer to the incubation buffer and boiling at 100°C for 5 min. The neuroaminidase can also deglycosylates the α–dystroglycan but no direct action of MMPs on this glycoprotein has been reported previously. In our experiments we have not examined the impact of the processing on α-dystroglycan glycosylation.

Limited proteolysis treatment with a specific protease solution (1mg/ml of fresh trypsin solution) of the protein Schwann cells homogenate containing Up395, Up71, and β-DG (43 and 30 kDa) was performed to test the behaviour of these proteins toward such conditions.

**Immunoprecipitation**

Immunoprecipitation was performed using 1) the supernatant containing the two isoforms of β-DG and 2) the crude membrane fraction after enzymatic deglycosylation. The samples were incubated with the specific antibody (NCL-b-DG, DYS2, and DRP1) overnight at 4°C with gentle agitation. The protein-antibody complex was mixed with magnetic protein A micro-beads (MACS: Miltenyi Biotec). After extensive washing with a buffer containing 0.1% Triton X-100, 50 mM Tris/HCl, pH 8, 120 mM NaCl, 0.75 mM bezamedine and 0.1 mM PMSF, the magnetically labelled immune complex was purified over a micro-column placed in the magnetic field of the MACS separator. The bound fractions were eluted by a buffer containing 0.1% Triton X-100, 50 mM Tris/HCl, pH 7.4, 0.35 Nacetyl-D-glucosamine, 0.75 mM bezamedine and 0.1 mM PMSF. Protein concentrations were estimated using the BCA protein assay (Pierce) to equilibrate samples. The experiments were performed in duplicate using total PN homogenate and Schwann cells crude membrane of C57BL/10 and mdx mice.

**Blot overlay assay**

Proteins were separated by 5-15% SDS-PAGE and transferred to nitrocellulose membrane. The nitrocellulose transfers were blocked for 1 h at room temperature with 10 mM triethanolamine, pH 7.6, 140 mM NaCl, 1mM CaCl2, and 1mM MgCl2 containing 5% BSA (bovine serum albumin) and then incubated with the same solution containing 1mM dithiothreitol and the crude membrane cell homogenate overnight (O/N) at 4°C with gentle agitation (Imamura et al. 2000). After washing, the dystrophin and utrophin isoforms, which had bound to β-DG on the nitrocellulose transfers, were detected by C-ter dystrophin or C-ter utrophin antibodies.

**Semi-quantitative RT-PCR**

Total RNA was isolated from $10^8$ Schwann cells with an SV total RNA isolation kit (Promega) following the manufacturer's instructions. We used 0.1 µg of total RNA to generate first strand cDNA sequences with random hexanucleotide primers and the M-MLV reverse transcriptase (invitrogen). Of the RT reaction, 12 µl were subsequently used for each PCR reaction in 25 µl using the Taq Polymerase from Qbiogene. Three PCR reactions were performed to amplify the mouse Up395 isoform using the primers (mUp-ex17/20-F and mUp-ex17/20-F) previously described (Wilson et al. 1999). The mouse Up71 used a forward primer located in the unique first exon (FUP71, 5'-TTGAATACTGAGTAATAATTGAGTACTAG-3'), and a reverse primer located in exon 63 (RUp71, 5'-AAGAGCTCATTTTAGGATGAT-3'). The mouse β-DG gene fragment was amplified using the following primers: Fdag1-5'-GGAGGCTGTTCCCACCGTGGT-3'; Rdag1, 5'-CTCTGCATTCTGTTCAACAGATCG-3'. The obtained PCR products were 383-pb (mUp395), 220-pb (mUp71) and 474-pb (dag1), respectively. The PCR conditions included 94°C for 5 min, 35 cycles of 94°C for 30 s, 60°C for 30 s, 72°C for 1 min, and a final extension of 72°C for 8 min. Assays was performed in duplicate and the different mRNA was expressed as a fraction of the House-keeping gene GAPDH which does not change between normal and mdx

Schwann cells. cDNA GAPDH amplification was used identically as the other PCR conditions to normalise expression level of Up395, Up71, and β-DG mRNA as described by Demestre et al. (Demestre et al.2004). Negative controls were performed, with total RNA replaced by either RNAase free water or reaction mixtures without RT (for DNA contamination). PCR products were visualized on 1.5% agarose gel. The 100-pb molecular mass markers (Promega) were used to estimate the molecular mass of the PCR products.

**Statistical analysis**

Immunolabeled sections were analyzed by Histolab program as described above. For each measure, images were taken under identical conditions and at the same objective. Data were averaged per dissected PN obtained from 3 C57BL/10 or mdx mice and statically compared using the man-Whitney test. Displayed data (mean ±S.D.) were normalized to the staining intensity of wild-type and comparisons were considered significant when *$P$<0.05. For Zymograhy and PCR band analysis scanned gels were treated in black/white colour and analysed by the NIH Image software package. Data were compared between normal and mdx mice using man-Whiney test as described above.

**RESULTS**

**Peripheral nerve localization of Dp116, utrophin and associated proteins in C57BL/10 and mdx mice**

Immunofluorescence analysis of transversal cryostat sections of C57BL/10 and mdx nerve was performed using antibodies raised against Dp116, utrophin, and β-DG (Fig. 1). We observed identical labelling of the short dystrophin isoform (Dp116) in normal and dystrophin-deficient mice. The utrophin and β-DG staining was relatively higher in mdx PN sections (Fig. 1A and 1B). This result was confirmed by fluorescence intensity analysis of several sections of C57BL/10 and mdx peripheral nerve. Furthermore, using dissected PN to visualize nodes of Ranvier, Utrophin and β-DG labelling was stronger in several mdx nodes in comparison to C57BL/10 nodes (Fig. 2A). This is approved by the comparative analysis of fluorescence intensity using several nodes of Ranvier (20 nodes) obtained from each mdx and C57BL/10 dissected PN (Fig. 2B). Moreover, double labelling of Schwann cells with S-100 antibody and utrophin or β-DG antibodies showed also a high expression pattern of these tow proteins (Utrophin and β-DG) in the Schwann cells outer membrane of mdx nerve (Fig. 2C, 2D).

**Characterization of the short utrophin isoform Up71 in mdx mouse peripheral nerve**

In an attempt to establish a correlation between the intensive labelling of utrophin in the outer region of nerve fibres and its expression level, we analyzed total PN extracts of C57BL/10 and mdx mice (Fig. 3A). We detected a protein band with a Mr. 70 kDa of the utrophin family products, previously described by our team and others (Fabbrizio et al. 1993; Rivier et al. 1997; Wilson et al. 1999) in mdx PN extract by using three antibodies which recognize, respectively, both utrophin and dystrophin (Fig. 3A, panel a), only C-terminal dystrophin (Fig 3A, panel b) and only the C-terminal end of utrophin (Fig. 3A, panel c). We assumed that this protein was the short product Up71 of the utrophin gene (Wilson et al. 1999) because our antibody also recognized the same C-terminal sequence of the full-length utrophin (Up395). Western blot analysis revealed that Up71 was highly detected in mdx nerve. To further confirm our results, we performed semi-quantitative RT-PCR analyses of Up71, β-DG and Up395 mRNA in C57BL/10 and mdx Schwann cells using specific primers for each transcript (Fig. 3B). When normalized to the GAPDH fraction Up71 mRNA level was strongly detected in mdx samples but weakly detected in C57BL10 samples. Up395 and β-DG had identical expression in normal and mdx samples.

**Post-translational processing of β-DG in C57BL/10 and mdx nerve**

Western blot detection of β-DG and the other associated proteins in total PN extract showed that the two isoforms of this glycoprotein were present (Fig. 3C, panel a). It was reported that a 30 kDa β-DG is detected in peripheral nerve extract and that this fragment is the product of proteolytic processing of the extracellular domain of the β-DG$_{full}$ by matrix metalloproteinase (MMP) activity (Yamada et al. 2001). Our results confirmed this observation in both C57BL/10 and mdx PN. Interestingly, the proteolytic processing in mdx PN seemed to be more highly activated, leading to a greater amount of the β-DG$_{30}$, based on our Western blot detection (Fig. 3C, panel a). In parallel, no difference was observed in the α-DB1 and α7B-integrin levels between C57BL/10 and mdx PN extracts (Fig 2C, panel b and c, respectively).

**MMP in normal and mdx PN**

Zymographic analysis of normal (C57BL/10) nerve extracts showed MMP-2 (major 60 kDa and minor 66 kDa latent forms), whereas both MMP-2 and a small amount of MMP-9 (100 kDa) were detected under the same non-reducing conditions in mdx nerve (Fig. 4A). MMP-2 and MMP-9 were produced as zymogens in normal mouse peripheral nerves, with MMP-2 being the major form present in these nerves. It is constitutively produced by Schwann cells and neurons (Muir 1995; Kherif et al. 1998). In an additional experiment we demonstrate that time course treatment with a protease solution (1 mg/ml trypsin) of Schwann cell homogenate containing Up395, Up71, and β-DG isoforms fail to produce any utrophin product with Mr. corresponding to Up71 or β-DG product with a Mr. of 30 kDa detectable by the corresponding antibodies in both C57BL/10 and mdx homogenate (Fig. 4B). To confirm the specific action of MMPs on β-DG$_{full}$ into processing β-DG$_{30}$, Scwann cell was treated for 3 days by the MMPs inhibitor, *N*-biphenyl sulfonyl-phenylalanine hydroxiamic acid (BPHA) (Maekawa et al. 1999), as described by Yamada et al.

(Yamada et al. 2001). Cells were harvested and analysed by western blot using β-DG C-ter antibody (Fig. 4C).

**Peripheral nerve β-DG$_{full}$ and β-DG$_{30}$ form two complexes that respectively anchor the Dp116 and utrophin isoforms**

All proteins detected in the peripheral nerve membrane were located along the outer membrane of the Schwann cells. To further investigate whether Up71 interacts with the transmembrane β-DG, we analyzed the mdx Schwann cell crude membrane preparation by immunoprecipitation. Using the monoclonal β-DG antibody, utrophin and Up71 were immunoprecipitated from the crude membrane fraction in both normal and mdx Schwann cells and PN homogenate (Fig. 5A). In accordance with the Western blot results, the amount of precipitated Up71 relative to Up395 was higher in the mdx fraction than in the normal C57BL/10 fraction (Fig. 5A, panel i). In both normal and mdx fraction, Dp116 was immunoprecipitated in equal amounts (Fig 5A, panel ii). Like Dp116, utrophin and the predicted sequence of Up71 had the same dystroglycan-binding site and were shown to bind to the same region of the β-DG cytoplasmic domain (Wilson et al. 1999). This suggests that these proteins bind to β-DG in a competitive manner. Immunoprecipitation using the anti-Dp116 antibody (DYS2) coimmunoprecipitated β-DG$_{full}$ and β-DG$_{30}$ in the Schwann cells of both C57BL/10 and mdx crude fraction membrane (Fig. 5B, panel i). On the other hand, with immunoprecipitation assay using the C-ter utrophin antibody (DRP1), we detected more β-DG$_{30}$ in the mdx fractions (Fig 5B, panel ii). This result is in agreement with previous reports which suggested that the interaction of utrophin with β-DGfull is weaker than that of Dp116 (Imamura et al. 2000). Another finding was that utrophin isoforms have more affinity to β-DG$_{30}$, suggesting that processing of the β-DG may influence interactions between the other DAP complexes in Schwann cell membrane.

**Effect of β-DG processing on Dp116 and utrophin isoform anchorage**

Purified crude fraction containing β-DG$_{30}$ and β-DG$_{full}$ in mdx Schwann cells was treated with neuraminidase to obtain the unique β-DG$_{30}$ form (Fig. 6a). Dp116 and the utrophin isoforms were immunoprecipitated using the β-DG antibody (Fig. 6b and 6c, respectively). We observed that the immunoprecipitated Dp116 was more abundant in the untreated neuraminidase samples (Fig. 6b), whereas Up71 was abundantly immunoprecipitated in the treated samples (Fig. 6c). The utrophin full length Up395 was weakly immunoprecipitated in both treated and untreated samples (Fig. 6c). Dp116 seemed to have more affinity to β-DG$_{full}$, whereas β-DG$_{30}$ anchored the utrophin isoforms (Up395 and Up71). This was more evident in mdx Schwann cells, depending on the relative abundance of the β-DG$_{30}$ form, which resulted from the activation of the processing. To confirm this result, mdx Schwann cell crude membrane was separated on SDS-PAGE, transferred to nitrocellulose membranes, and incubated with the elute fraction of the IP-β-DG assay containing Dp116 and utrophin isoforms for overlay assay (Fig. 7a and 7b, respectively). After overlay, nitrocellulose membrane was revealed with dystrophin and utrophin antibodies, respectively. Dp116 seems to associate more with β-DG$_{full}$ band (Fig. 7a), whereas the utrophin isoforms (especially Up71) have more affinity with β-DG$_{30}$ band (Fig. 7b).

**DISCUSSION**

Taken together, the data presented here indicate two types of dystroglycan interaction in normal and mdx Schwann cells. The dystroglycan complex anchors Dp116, the short dystrophin isoform, and the utrophin isoforms. This interaction depends of the endogenous processing of β-DG, the transmembrane glycoprotein, by matrix metallonoproteinases (MMPs) (Yamada et al. 2001). MMPs have essentially two active forms (MMP-2 and MMP-9) and their activity is responsible for processing β-DG$_{full}$ (43 kDa) into β-DG$_{30}$ (30 kDa). The cleavage site was predicted to be in the extracellular domain of β-DG, and β-DG$_{30}$ was predicted to be its C-terminal fragment. $\beta$-DG$_{30}$ is increased in the skeletal and cardiac muscles of cardiomyopathic hamsters, the model animals of sarcoglycanopathy (SGCP), and in the skeletal muscle of DMD patients, but not in other muscular diseases (Matsumura et al. 2005). Here, we investigated the processing of $\beta$-DG in normal (C57BL/10) and mdx nerve and its impact on Dp116/utrophin anchorage in Schwann cell membrane. In normal nerves, MMP-2 was strongly detected while MMP-9 was weakly expressed. Both pro- and active forms of MMP-2 and MMP-9 are expressed in mdx muscle, whereas only MMP-2 is constitutive in normal muscle (Kherif et al. 1999). This agrees with the idea that the combined activity of MMP-2 and MMP-9 in mdx and DMD muscle extracts could explain the greater amount of $\beta$-DG$_{30}$ in these extracts than in C57BL/10 extracts (Yamada et al. 2001; Roma et al. 2004). MMP-9 is activated in crushed nerve and in mdx muscle (Kherif et al. 1998; Kherif et al. 1999). As MMP-9 can degrade myelin basic protein (Yang and Bryant 1994; Platt et al. 2003), an increase in proteolytic activity due to the release of MMP-9 by macrophages may enhance myelin degradation (Kieseier et al. 1999). Proteolysis may also be required to free Schwann cells from their basement membrane connection as they proliferate and to re-establish axonal connections (Ferguson and Muir 2000). This finding leads us to suggest that the greater amount of $\beta$-DG$_{30}$ observed in mdx PN extract was probably due to an extrinsic activation of MMP-9 by infiltrating macrophages within the mdx dystrophic muscle (Kherif et al. 1999). In fact, it was shown that in mdx muscle MMP-9 expression is related to the inflammatory response and probably to the activation of satellite cells, whereas MMP-2 activation is concomitant with the regeneration of new myofibers (Barani et al. 2003). In our experiments, we showed that mdx nerve expresses more MMP-9 than normal nerve and such activation correlates with the increased level of $\beta$-DG$_{30}$. In a second part of the experiment, we investigated the expression of Dp116 and utrophin comparatively in mdx and normal PN. Western blot analysis revealed that the utrophin short isoform (Up71) was detected in mdx extracts. We characterized this isoform by RT-PCR analysis and Western blot detection using a C-terminal antibody directed against the last ten amino acids of the utrophin sequence. Several groups have observed small gene products using antibodies raised against the utrophin C-terminus (Fabbrizio et al. 1995; Rivier et al. 1996). Up71 is a short transcript of utrophin with a unique first exon located in intron 62 and predicted to encode a 4.0 kb mRNA and a 71 kDa protein with the same cysteine-rich and C-terminal domains as full-length utrophin, commencing with exon 63 (dystrophin exon numbering) (Wilson et al. 1999). These domains are implicated in interactions with β-DG (Tinsley et al. 1992). Like full-length utrophin, the Up71 isoform is ubiquitously expressed and was detected by our team and others in the peripheral nervous system and the small peripheral arteries (Fabbrizio et al. 1995; Rivier et al. 1996, 1997; Wilson et al. 1999). To date there has been little information about the expression of this short transcript and its implication in DMD pathology. We demonstrate here that Up71 expression is increased in mdx nerve. Immunoprecipitation and blot overlay assay showed that Up71 has more affinity to β-$DG_{30}$ than the full-length isoform Up395, while Dp116 interacts with β-$DG_{full}$ in both C57BL/10 and mdx Schwann cell membrane. This result agrees with the findings of Imamura et al., who suggested that the interaction of Up395 with β-$DG_{full}$ is weaker than that of Dp116 (Imamura et al. 2000). In agreement with previous reports, we showed that β-$DG_{full}$ and β-$DG_{30}$ have identical extraction profiles (Yamada et al. 2001). They were both extracted by 2% Triton X-100 and high ionic strength enhanced the effect of the detergent. Because β-DG is a type I integral membrane protein with a single transmembrane domain, these results indicate that β-$DG_{30}$ retains this transmembrane domain. This is further supported by the finding that β-$DG_{30}$ is recognized by the C-terminal antibody of β-$DG_{full}$. What effects this processing have on DAP complex integrity in Schwann cell membrane? The finding that β-$DG_{30}$ did not interact with α-dystroglycan, indicates that the cleavage of β-$DG_{full}$ into β-$DG_{30}$ disintegrates the dystroglycan complex. Because α- and β-DG are responsible for binding to the basement membranes, this cleavage will disrupt the link between the basement and Schwann cell membrane. Proteolysis mediated by MMPs may also be required to free the Schwann cells from their basement membrane connection as they proliferate. This could explain why no pathological effect was noted in mdx nerve despite the constant MMP activation and $\beta$-DG$_{30}$ accumulation. However, MMPs were active in mdx skeletal muscle and, during the degeneration-regeneration process; MMP-9 is essentially produced by both inflammatory and activated satellite cells (Guerin and Holland 1995). These cells produce MMP-9 in response to different stimuli from necrotic tissue and are an important source of MMP-9. It is well established that macrophages from blood infiltrate the injured nerve and are responsible for clearing myelin and other debris (Kherif et al. 1998). We suggest that the spatially and temporally limited MMP-9 activity in mdx nerve is mediated by inflammatory and degenerative-regenerative processes in the microenvironment of necrotic mdx fibres. On the other hand, the increase in MMP-9 may help break down the blood-nerve barrier and thus facilitate macrophage infiltration. This activation is not sufficient to promote disruption of mdx nerve function. Our results suggest that $\beta$-DG processing is a constitutive process in normal and mdx nerves. The persistence and the high level of $\beta$-DG$_{30}$ in mdx Schwann cell membrane could be explained by its specific anchorage to the short utrophin isoform Up71 which confer to $\beta$-DG$_{30}$ more stabilization in membrane. This could be an argument for a potential implication of these two proteins in the degenerative-regenerative process of mdx nerve (See the hypothetical model proposed in Fig. 8). This idea is all the more attractive because $\beta$-DG interacts with adaptor proteins like Grb2 and rapsyn, which are implicated in signalling cascades during nerve regeneration. Our data suggest a novel role for $\beta$-DG$_{30}$ and the short utrophin transcript Up71 in the Schwann cell membrane and nerve physiology.

Acknowledgments: This work was supported by the "Association Française contre les Myopathies" (AFM, Fellowship No. 10529), INSERM and CNRS.

**Figure Legends**

**Fig. 1.** (**A**) Localization of DAP complex in normal (C57B/10) and mdx transversal peripheral nerve sections. Dp116 (A1, A2) and utrophin (UTR) (A3, A4) were detected in Schwann cell outer membrane as shown by the S-100 antibody (A5, A6) which labels myelinated and non-myelinated Schwann cells. (**B**) β-dystroglycan (β-DG) (B1, B2), α-dystrobrevin1 (α-DB1) (B3, B4), and α7B-integrin (α7B-ITG) (B5, B6) detection on transversal PN section in C57BL/10 and mdx nerve showed different localization of these proteins in nerve. β-DG which showed the most intensive labelling and α-DB1 were localized in the in Schwann cell outer membrane whereas α7B-ITG is expressed in axon region. Bar = 15 µm.

**Fig. 2.** (**A**) Localization of Dp116 (A1, A2), Utrophin (UTR) (A5, A6) and β-dystroglycan (β-DG) (A9, A10) in nodes of Ranvier. A3, A7, A11 showed the specific labelling of Nodes of Ranvier (arrows) with the pan-Na$^+$ antibody and the Hematoxylin/eosin (H&E) staining (A4, A8, and A12). Dp116 seems deceased in some nodes of Ranvier in mdx sections (A2), whereas utrophin and β-DG (A6 and A10 respectively) were highly detected compared with C57BL/10 (A5 and A9). (**B**) Immunolabelling of utrophin (UTR) (B1, B3) and β-dystroglycan (β-DG) (B5, B7) in the outer membrane of Schwann cells on isolated C57BL/10 and mdx nerve fibres. The same Schwann cells (Arrow head) were also detected by the S-100 antibody (B2, B6 and B4, B8) 5µm.

**Fig. 3**. (**A**) Immunodetection of DAP complex in C57BL/10 and mdx peripheral nerve extracts. a) Western blots revealed a dystrophin isoform (Dp116) and utrophin isoforms (Up395 and Up71) in C57BL/10 and mdx total nerve extracts, using a monoclonal antibody recognizing both dystrophin and utrophin (5A3). b) Dp116 detection on the same extracts with C-terminal dystrophin antibody (H4). c) C-ter utrophin antibody (K7) detection showed the full utrophin (Up395) in C57BL/10 and both Up395 and the short product Up71 (71 kDa) in mdx extract. d) Represents Western blot detection of β-dystroglycan (β-DG) in C57BL/10 and mdx nerve

extracts. We noted that two isoforms of β-DG (43 kDa and 30 kDa) were present in PN extracts the 43 kDa isoform (β-DG$_{full}$) was more abundant in C57BL/10 extract than in mdx one. In the opposite, the 30 kDa (β-DG$_{30}$) was more represented in mdx PN extract than in normal one. All the samples were equilibrated according to β-actin detection. (**B**) Semi-quantitative RT-PCR analysis of a) Up395, b) β-DG and c) Up71 transcripts. PCR products after cDNA amplification of Up71, Up395 and β-DG of C57BL/10 and mdx Schwann cells nerves. Assay was performed in duplicate and the different mRNA was expressed as a fraction of the House-keeping gene GAPDH. The 100-pb molecular mass markers (Promega) were used to estimate the molecular mass of the PCR products. d) Statistical analysis of the mRNA level of the Up395, β-GD and Up71. Black histograms correspond to mdx samples whereas the white histograms represents that of wild-type (C75BL/10) samples.

**Fig. 4**. (**A**) Zymographic analysis of normal (C57BL/10) and mdx nerve. In normal nerve extract, both MMP-2 (major 60 kDa and minor 66 kDa latent forms) and MMP-9 (100 kDa) were detected under non-reducing conditions. In mdx extracts, both gelatinases were also expressed with a significant difference in the intensity of the MMP-9 band, which was more expressed compared with C57BL/10 extract. Histograms under the gels represent statistical analysis of the gelatinase bands in the two samples. Experiments were performed in triplicate and statistical significance (*) was set at $P < 0.05$. Data are expressed as mean ± SEM. (**B**) Limited proteolysis (0-35 second) with a fresh trypsin solution (1mg/ml, final concentration of 1%) showed disappearance of the endogenous Up395, Up71, and the two isoforms of β-DG (β-DG$_{full}$ and β-DG$_{30}$) in C57BL/10 (a) or mdx (b) and Schwann cell homogenates when detected with the C-ter utrophin antibody (left panels) or with C-ter β-DG antibody (right panel). Suppression of β-DG$_{full}$ into β-DG$_{30}$ was performed by MMP inhibitor (BPHA). Mdx Schwann cells were treated with 0.5μg/ml

or 20µg/ml BPHA and after 3 days cells were analysed by immunoblotting using β-DG C-ter antibody.

**Fig.5.** Co-mmunoprecipitation of β-dystroglycan, Dp116 and utrophin isoforms in Schwann cell membrane (SC) and total PN homogenate (PN). (**A**) Each membrane fraction of the normal and mdx Schwann cell and PN in both C57BL/10 and mdx mice was used for immunoprecipitation (IP) with a monoclonal antibody against β-DG (IP- β-DG). The protein composition of each precipitate was identified by immunoblotting with i) utrophin antibody and ii) dystrophin antibody. (iii) western blot of the endogenous β-DG in C57BL/10 and mdx homogenates. (**B**) The same membrane fractions were used for immunoprecipitation with i) monoclonal antibody against dystrophin (IP-DYS) and ii) utrophin (IP-UTR). Each precipitate was analyzed by immunoblotting using β-DG antibody that recognizes the two isoforms of β-DG (β-DG$_{full}$ and β-DG$_{30}$) identically in Schwann cells crude membrane and PN homogenate (Not shown).

**Fig.6**. Effect of neuraminidase treatment on the binding of Dp116 and utrophin isoforms to mdx peripheral nerve β-dystroglycan. a) Membrane fractions of mdx Schwann cells were incubated with or without neuraminidase (+ and -, respectively) and separated by 5-15% SDS-PAGE, transferred to nitrocellulose membrane and then revealed with β-DG antibody. After neuraminidase treatment, only the 30 kDa form of β-DG (β-DG$_{30}$) was present in the membrane fraction. Untreated and treated samples were used for immunoprecipitation with β-DG antibody (panel b and c). The untreated neuraminidase fraction contained both isoforms of β-DG (Neur–), while the treated fraction contained only the 30 kDa β-DG (Neur+). Each precipitate was electroblotted and reveled with C-terminal dystrophin (panel b) and C-terminal utrophin (panel c) antibodies.

**Fig.7**. Blot overlay analysis of the binding of Dp116 and utrophin isoforms with the Schwann cell membrane protein fraction of mdx mouse. Mdx Schwann cell membrane was separated by 5-15% SDS-PAGE, transferred to nitrocellulose membrane, and overlaid O/N at 4°C with the crude membrane fraction after β-DG immunoprecipitation (IP- β-DG) (see Fig. 5). After intensive washing, the nitrocellulose membrane was revealed by the dystrophin antibody (panel **a**) and the utrophin antibody (panel b). In panel a, the arrows show positive bands with an Mr. of 117 kDa (corresponding to the Dp116 band) and an intensive 49 kDa band revealed with dystrophin antibody that corresponds to the Dp116 fraction associated with the 43 kDa β-DG after overlay. Panel **b** shows the nitrocellulose membrane revealed by utrophin antibody after overlay. Three bands were detected (70 kDa, 49 kDa and 30 kDa). The 70 kDa band corresponds to Up71, the second band corresponds to the amount of utrophin fraction which bound the β-$DG_{full}$, and the last band (most intensive) corresponds to the utrophin fraction which bound to β-$DG_{30}$. Panel c represents the Western blot revelation of β-DG, utrophin, and dystrophin before overlay.

**Fig.8.** Hypothetical model of β-BG-Dp116, -Up395 and -Up71 interaction in C57BL/10 and mdx peripheral nerve. **a)** β-BG is localized in the Schwann cell outer membrane, whereas Dp116 and Up395 are localized in the Schwann cell cytoplasm. **b)** Normal (C57BL/10) and mdx Schwann cell membrane. The processing of β-$BG_{full}$ to β-$BG_{30}$ by MMPs (MMP-2 and MMP-9) leads to the disruption of the link between the basal lamina and the outer membrane. The Up71 anchorage to β-$BG_{30}$ could stabilize this isoform in Schwann cell membrane to avoid any dysfunction in mdx nerve physiology.